%Paper: hep-th/9304051
%From: fendley@BUCRF17.BU.EDU (Paul Fendley)
%Date: Tue, 13 Apr 93 13:09:41 -0400

% This paper uses harvmac macros
\input harvmac
\def\t{\theta}\def\tht{\theta}
\def\g{\mu}
\def\l{\lambda}
\def\cc{\vrule width.3cm height .4pt}
\let\p=\prime
\def\e{\epsilon}
\lref\Zamocfn{A.B. Zamolodchikov, JETP Lett. 43 (1986) 730.}
\lref\LC{A. Ludwig and J. Cardy, Nucl. Phys. B285 (1987) 687.}
\lref\RSOS{A.B. Zamolodchikov, Landau Institute preprint, September 1989.}
\lref\Leclair{ A. LeClair, Phys. Lett. 230B (1989) 103.}
\lref\ZamoIII{Al.B. Zamolodchikov, Nucl. Phys. B358 (1991) 524.}
\lref\Gri{M.T. Grisaru, A. Lerda, S. Penati and D. Zanon,
 Phys. Lett. B234 (1990) 88; Nucl. Phys. B342 (1990) 564.}
\lref\ZamoII{Al.B. Zamolodchikov, Nucl. Phys. B358 (1991) 497.}
\lref\ZamoI{Al.B. Zamolodchikov, Nucl. Phys. B342 (1990) 695.}
\lref\Bernard{D. Bernard, Phys. Lett. B279 (1992) 78.}
\lref\Baxter{R. Baxter, {\it
Exactly Solved Models in Statistical Mechanics} (Academic Press, 1982).}
\lref\rTS{M. Takahashi and M. Suzuki, Prog. Th. Phys. 48 (1972) 2187.}
\lref\FSI{P. Fendley, H. Saleur and Al.B. Zamolodchikov,
``Massless Flows I: the sine-Gordon and $O(n)$ models'', BUHEP-93-5,
USC/93-003, LPM-93-07.}
\lref\Huse{D.A. Huse, Phys. Rev. B30 (1984) 3908.}
\lref\FS{P. Fendley and H. Saleur, Nucl. Phys. B388 (1992) 609.}
\lref\ABF{G. Andrews, R. Baxter and J. Forrester, J. Stat. Phys. 35
(1984) 193.}
\lref\Resh{N. Yu. Reshetikhin, J. Phys. A24 (1991) 3299.}
\lref\Pasquier{V.Pasquier, Comm. Math. Phys. 118 (1988) 355.}
\lref\KM{T. Klassen and E. Melzer, Nucl. Phys. B370 (1992) 511.}
\lref\ZamZam{A.B. Zamolodchikov and Al.B.  Zamolodchikov, Nucl. Phys. B379
(1992) 602.}
\lref\Zamopol{A.B. Zamolodchikov, Mod. Phys. Lett. A6 (1991) 1807.}
\lref\Smir{F.A. Smirnov, Phys. Lett. B275 (1992) 109.}
\def\wgta#1#2#3#4{\hbox{\rlap{\lower.35cm\hbox{$#1$}}
\hskip.2cm\rlap{\raise.25cm\hbox{$#2$}}
\rlap{\vrule width1.3cm height.4pt}
\hskip.55cm\rlap{\lower.6cm\hbox{\vrule width.4pt height1.2cm}}
\hskip.15cm
\rlap{\raise.25cm\hbox{$#3$}}\hskip.25cm\lower.35cm\hbox{$#4$}\hskip.6cm}}
\def\wgtb#1#2#3#4{\hbox{\rlap{\raise.25cm\hbox{$#2$}}
\hskip.2cm\rlap{\lower.35cm\hbox{$#1$}}
\rlap{\vrule width1.3cm height.4pt}
\hskip.55cm\rlap{\lower.6cm\hbox{\vrule width.4pt height1.2cm}}
\hskip.15cm
\rlap{\lower.35cm\hbox{$#4$}}\hskip.25cm\raise.25cm\hbox{$#3$}\hskip.6cm}}
\def\wgtc#1#2#3#4{\hbox{\rlap{\vrule width1.3cm height.4pt}
\rlap{\lower.35cm\hbox{$#1$}}\rlap{\raise.25cm\hbox{$#2$}}
\hskip.38cm\rlap{\lower.6cm\hbox{\vrule width.4pt height1.2cm}}
\hskip.22cm
\rlap{\lower.35cm\hbox{$#4$}}\raise.25cm\hbox{$#3$}\hskip.3cm}}

\Title{\vbox{\baselineskip12pt\hbox{BUHEP-93-6, USC-93/004, LPM-93-08}}}
{\vbox{\centerline{Massless Flows II:} \vskip4pt\centerline{the exact
$S$-matrix approach}}}

\centerline{P. Fendley$^\#$, H. Saleur$^\dagger$ and Al. B. Zamolodchikov$^*$}
\medskip\bigskip\centerline{$^\#$Department of Physics, Boston University}
\centerline{590 Commonwealth Avenue, Boston, MA 02215}
\bigskip\centerline{$^\dagger$Department of Physics,
University of Southern California}
\centerline{Los Angeles, CA 90089}
\bigskip\centerline{$^*$Laboratoire de Physique Math\'ematique}
\centerline{Universit\'e Montpellier 2, Montpellier, France}
\bigskip

\vskip .2in We study the spectrum, the massless $S$-matrices and the
ground-state energy of the flows between successive minimal models of
conformal field theory, and within the sine-Gordon model with imaginary
coefficient of the cosine term (related to the minimal models by
``truncation'').

For the minimal models, we find exact $S$-matrices which describe the
scattering of massless kinks, and show using the thermodynamic Bethe ansatz
that the resulting non-perturbative $c$-function (defined by the Casimir
energy on a cylinder) flows appropriately between the two theories, as
conjectured earlier.

For the non-unitary sine-Gordon model, we find unusual behavior.  For the
range of couplings we can study analytically, the natural $S$-matrix deduced
from the minimal one by ``undoing'' the quantum-group truncation does not
reproduce the proper $c$-function with the TBA. It does, however, describe the
correct properties of the model in a magnetic field.

\bigskip
\Date{April 1993}

\newsec{Introduction}

One of the most useful methods of understanding an integrable
two-dimen\-sional field theory off the critical point has been to find the
exact $S$-matrix. Strict consistency criteria have allowed exact $S$-matrices
for a large number of massive integrable models to be conjectured, using
intuition from perturbed conformal field theory. Many have been essentially
verified by a number of checks, especially that of the thermodynamic Bethe
ansatz, where one can calculate the Casimir energy from the $S$-matrix and
check that it is correct at the conformal point.

Massive theories are not the only type of field theory which can be studied by
these methods. There are numerous situations where perturbation of a conformal
theory by a relevant operator causes it to flow to another conformal theory.
In the midst of the flow, the model has massless excitations (the only ones to
survive in the infrared limit) but is not scale-invariant. Thus one can hope
to find an exact $S$-matrix for these excitations, and calculate the associated
Casimir energy in order to study the flow. This has been done for several
models in \refs{\ZamoIII,\ZamZam}. There are subtleties in defining an
$S$-matrix for massless particles, but since our models are integrable and have
an infinite number of conserved currents, we expect that the idea of
massless-particle scattering can be implemented. In particular, it was shown
some time ago that in the lattice XXX model (which is at a conformal point),
one can define an $S$-matrix by deriving phase shifts of quasi-particle
excitations from the Bethe ansatz \ref\XXX{L.D Faddeev, L.A. Takhtajan, Phys.
Lett. A85 (1981) 375;\hfill\break N. Andrei, K. Furuya and J. Lowenstein, Rev.
Mod. Phys. 55 (1983) 331}.

One crucial point to realize about a massless $S$-matrix is that it should be
viewed as describing an (irrelevant) perturbation of the {\bf infrared} fixed
point.  The reason for this is simple: the massless excitations are those
which remain in the spectrum at the infrared fixed point. All known massless
$S$-matrices follow this pattern; one mainly uses intuition gained by studying
the infrared, not the ultraviolet, conformal field theory.  We can even make
this observation precise.  In these massless flows, the mass $M$ provides the
scale; $M=0$ gives the ultraviolet fixed point while $M\rightarrow\infty$
gives the infared one. Since the particles are massless, they are either
right- or left-moving; we parametrize the left-movers' momenta by
$p_L=-(M/2)\exp (-\t)$, and the right-movers by $p_R=(M/2)\exp \t$.  A
Lorentz-invariant $S$-matrix element $S_{LL}$ describing scattering of two
left movers depends only on the ratio of the two momenta, so it depends only
on $\t_1-\t_2$ and not on $M$. We can always rescale $M\rightarrow\infty$ by
shifting the rapidities. The $LL$ and $RR$ $S$-matrices are independent of
this shift (although $S_{LR}$ is not), so they are characterized solely by
properties of the infrared fixed point. In fact, one can think of the $LL$ and
$RR$ $S$-matrices as being the $S$-matrices for the conformal field theory.
The $LR$ $S$-matrix does depend on $M$, so this is affected by the
perturbation.

There are two purposes of this paper.  The first is to describe the flows
between the minimal models of conformal field theories, completing a picture
initiated in \refs{\ZamoII,\ZamoIII}.  We will show that the spectrum consists
of massless kinks, which can be thought of as interpolating between the vacua
of a scalar field with a degenerate-well potential.  Their $S$-matrix is
related to a well-known RSOS one. We will verify that these $S$-matrices give
the appropriate ground-state energy (on a cylinder) in the IR limit in all
cases, and for all scales in the simplest cases. The ground-state energy
provides a non-perturbative $c$-function which allows one to see explicitly
the crossover from one minimal model to another, a phenomena which had been
earlier studied perturbatively \refs{\Zamocfn,\LC}.

The second purpose is to discuss the related $S$-matrix simply obtained by
``undoing'' the quantum-group truncation, which may describe scattering in the
sine-Gordon model with imaginary coefficient of the cosine term. Besides its
relation with minimal models, this theory is interesting in its own right: it
is the continuum limit of the $O(n)$ model in the low temperature phase, and
for $n\rightarrow 0$, it describes the ``dense'' phase of self-avoiding
polymers.  As we have discussed at length in our previous paper \FSI, this
non-unitary sine-Gordon model has a number of interesting but
difficult-to-handle properties. Our discussion will further elucidate these
properties, although we will see that the $S$-matrix does not seem to
completely describe the model. It does reproduce the correct behaviour with a
magnetic field in infinite volume, but we will show (at least in the simplest
cases) that for properties on a cylinder (i.e.\ at finite temperature) the
behavior in the ultraviolet is incorrect.

Various scenarios to explain this feature are proposed.

\newsec{The flows between the minimal models}

\subsec{$S$-matrix}

The models we study in this section are the minimal models of conformal field
theory perturbed by their least-relevant operator $\Phi_{13}$:
\eqn\pert{S=S^*_{\hbox{min}}(t)+\delta\beta\int\Phi_{13}\ d^2 x,}
where the unperturbed conformal field theory has central charge
$c=1-6/t(t+1)$. The physical properties depend crucially on the sign of
$\delta\beta$. For $\delta\beta>0$, the model is massive.  The underlying
lattice model \ABF\ has $t$ degenerate vacua \Huse. An effective
Landau-Ginzburg theory \ref\ZamoLG{A.B.  Zamolodchikov, Sov. J. Nucl.  Phys.
44 (1986) 529}\ describes the theory by a single scalar field with a potential
with $t$ degenerate wells.  The particles are kinks which interpolate between
adjacent vacua, and an $S$-matrix consistent with the quantum-group symmetry
was conjectured in \refs{\RSOS,\Leclair}. This $S$-matrix was checked by
showing that the resulting exact Casimir energy gives the correct result at
the conformal point \refs{\ZamoII,\FS}.

For $\delta\beta<0$ the model is qualitatively different: it flows from the
$t$-th minimal conformal field theory to the $(t-1)$-st one
\refs{\Zamocfn,\LC}. In the midst of the flow, the model should have massless
excitations. In the $t=4$ case (the flow from the tricritical Ising model to
the Ising model) the spectrum was shown \ZamoIII\ to consist of a left-moving
and a right-moving particle, which can be thought of as the massless
Goldstinos arising from spontaneously-broken supersymmetry. Their $S$-matrix
was found, and the ground-state energy was calculated. The effective central
charge was shown to smoothly flow from $7/10$ in the ultraviolet to $1/2$ in
the infrared. In addition, the ground-state energy was also conjectured for
all $t$ without benefit of an $S$-matrix. This was not omniscience; looking at
the massive case for all $t$ and the massless case for $t=4$ makes the guess
straightforward.  This was shown in \refs{\ZamoIII,\KM} to have the correct
properties.

Obviously, it would be useful to find the $S$-matrix for all $t$, and to then
verify that the ground-state energy conjectured in \ZamoIII\ is correct. We
will find the $S$-matrix for all $t$, showing that it agrees with the
already-known $t=4$ case. We find that in the general case, the spectrum
consists of kinks reminiscent of the massive case, but here they are massless.
Although one is used to thinking of kinks as field configurations with
non-zero energy, there is no reason why they cannot be degenerate with the
ground state in a quantum theory.  We will explicitly calculate the
ground-state energy for $t=5$ and find the conjectured result.  Technical
complications prevent us from doing so for all $t$, but we show that in the
infrared limit our $S$-matrices give the correct answer.

We must first find the appropriate particle content. For guidance, we can
again study the underlying lattice model \ABF. As discussed in \Huse, whereas
the massive phase (regime III) has $t-1$ degenerate vacua, the critical point
described by the $t^{\hbox{th}}$ minimal conformal field theory separates this
from a phase with $t-2$ degenerate vacua (regime IV). The result can be seen
in the effective Landau-Ginzburg picture: for even $t$, the central vacuum
disappears, and for odd $t$ the two central vacua coalesce into one.  The
dimension of the perturbing operator is the same in both phases, and since
there is only one such operator in the conformal theory (the $\Phi_{13}$
operator), the two phases must correspond to perturbing by $\Phi_{13}$ with
different signs. So the clear guess is that as in the massive case, there are
kinks interpolating between adjacent vacua, but for a given $t$ there is one
less vacuum. In addition, the kinks here are massless, so for every pair of
adjacent vacua we have a right-moving and a left-moving kink.

We derive the simplest $S$-matrix for this spectrum. Since scattering is
diagonal in the $L$ and $R$ labels, we can consider the two-particle
$S$-matrices $S_{LR}$, $S_{RL}$ and $S_{LL}=S_{RR}$ separately.\foot{It may
appear absurd to define an $S$-matrix element $S_{LR}$ for a right-moving
particle starting to the right of a left-moving one.  However, it should be
interpreted as a matching condition on plane-wave eigenstates of the
Hamiltonian \ZamoI.} The mass scale $M\propto |\delta\beta|^{(t+1)/4}$. The
$S$-matrix depends only on the rapidity difference $\theta=\t_1-\t_2$: for
$LL$ scattering, the ratio of the two left-moving particles' momenta is the
only Lorentz-invariant kinematic quantity, and for $RL$ scattering the Lorentz
invariant $s=(p_1+p_2)^2$ also depends only on the rapidity difference.  In
order that the scattering be factorizable, the $S$-matrix must obey
Yang-Baxter equation for a three-particle state:
\eqn\YBE{S_{12}(\t_1-\t_2)S_{13}(\t_1-\t_3)S_{23}(\t_2-\t_3) =
S_{23}(\t_2-\t_3)S_{13}(\t_1-\t_3)S_{12}(\t_1-\t_2)}
where $S_{12}$ is the two-particle $S$-matrix acting on particles $1$ and $2$.
A two-kink configuration can be labeled by three vacua; a two-particle
$S$-matrix element can be labeled by four because only the middle vacuum can
change in a collision. The simplest possibility for an $LL$ or $RR$ $S$-matrix
obeying \YBE\ and having this structure is the well-known RSOS $S$-matrix \ABF
\eqn\LLRSOS{\eqalign{&\wgtb {m}{m\pm1}m{m\mp1}  = Z(\theta)
\left({\beta_m\over\beta_{m+1}^{1/2}\beta_{m-1}^{1/2}}\right)^{i{\t\over\pi}}
\sinh\g(i\pi-\t)\cr
&\wgta {m\pm1}m{m\mp1}m=Z(\t)\left(
{\beta_{m+1}^{1/2}\beta_{m-1}^{1/2}\over \beta_m}\right)^{1+i{\t\over\pi}}
-\sinh\g\t \cr
&\wgta {m+1}m{m+1}m=
Z(\t)\left({\beta_{m+1}\over \beta_m}\right)^{i{\t\over\pi}}
{\beta_1\over \beta_m}\sinh\g(im\pi+\t)\cr
&\wgta{m-1}m{m-1}m= Z(\t)
\left({\beta_{m-1}\over \beta_{m}}\right)^{i{\t\over\pi}}
{\beta_1\over \beta_m}\sinh\g(im\pi-\t)\cr}}
where
$$\beta_m=\sinh(im\g\pi)$$
and $\mu=1/(t-1)$. The horizontal line denotes a right-moving particle with
rapidity $\t_1$, while the vertical (moving from bottom to top) has rapidity
$\t_2$, and $\t=\t_1-\t_2$.  The allowed vacua run from $1$ to $t-2$, and
notice that as required, adjacent vacua differ only by $\pm 1$.  The factors
of $(\beta_m)^{i \t/\pi}$ ensure crossing symmetry, but can be absorbed by a
rapidity-dependent change of basis.  In the massive high-temperature phase the
scattering is given by the RSOS $S$-matrix \LLRSOS\ but with $\g=1/t$
\refs{\RSOS,\Leclair}. This of course is a manifestation of the fact that
there are $t-2$ vacua in the low-temperature phase, but $t-1$ at high
temperature.

In \ZamZam, two types of $RL$ $S$-matrices were discussed: one proportional to
the identity and the other $S_{LL}$. The same possibilities can be considered
in our case, since both trivially still obey $\YBE$, but they are not correct
here (the TBA gives the wrong central charge in the UV limit), probably
because they have too much symmetry.  There is a ``restricted quantum affine
symmetry'' \ref\BL{D. Bernard and A. Leclair, Nucl. Phys. B340 (1990) 721.}
which does exist in the critical limit but which is broken in the midst of the
flow.  Consider for instance the $t=5$ case and the flow from the three-state
Potts model to the tricritical Ising model. The restricted quantum affine
symmetry of the infrared fixed point, the tricritical Ising model, is in fact
$N$=1 supersymmetry. The flow approaches this fixed point by the irrelevant
operator $\Phi_{3,1}$, which in this case is the supersymmetry operator, of
dimension (3/2,3/2).  This operator is the lower component of a superfield
with the energy-momentum tensor. In order to perturb a model and preserve the
supersymmetry, one can only perturb by the upper component of a superfield.
Hence in this case the supersymmetry is explicitly broken off the critical
point.  As explained in the introduction, the $LL$ and $RR$ $S$-matrices
essentially describe only the infrared limit, so $S_{LL}$ and $S_{RR}$ should
have this symmetry (as they do: see \refs{\RSOS,\Leclair}), but $S_{RL}$
should break it. This excludes immediately the diagonal choice for $S_{RL}$.
The second choice $S_{RL}\propto S_{LL}$ was proposed in \Bernard\ where it
was argued to preserve ``diagonal supersymmetry''.  We are not sure what such
a concept means, but we have checked explicitly using the TBA that the
corresponding UV central charge is incorrect (we find $c_{UV}=9/5$).  Similar
arguments apply for all values of $t$ by replacing $N=1$ supersymmetry by the
appropriate restricted quantum affine symmetry. For the second choice, also
proposed in \Bernard, we find indeed incorrect values of $c_{UV}$ (except for
$t=4$ of course).  The two choices, $S_{RL}$ proportional to the identity or
to $S_{LL}$, therefore must be excluded.\foot{In fact, the $S_{RL}$ proposed
in \Bernard\ for the flows into the $SU(2)_k$ ($k>1$) WZW models do not give
the correct $c_{UV}=3$. The proposed $S_{LL}$ and $S_{RR}$ are correct (as
shown previously in \Resh), but $S_{RL}$ is equal to the one discussed below.}

There is fortunately a whole family of possible $RL$ $S$-matrices which obey
the Yang-Baxter equation \YBE. Consider the situation when one of the three
particles in \YBE\ is right-moving while the other two are left-moving. Here
one of the $S$-matrix elements in \YBE\ is of $LL$ type, while the other two
are of $RL$ or $LR$ type. If we shift the right-mover's rapidity from $\t_i$
to $\t_i- i\alpha$, \YBE\ is still of course obeyed. The key observation is
that the $S$-matrix elements affected by this shift are of $RL$ type if
$+\t_i$ appears in the argument and of $LR$ type if $-\t_i$ appears. The
equation is then equivalent to the all-$LL$ problem, and hence is solved, if
$$S_{RL}\propto S_{LL}(\t+i\alpha)\quad\qquad S_{LR}\propto
S_{LL}(\t-i\alpha).$$
To fix $\alpha$, we make the simple demand that $S_{LR}=S_{RL}$, because we
know that in the $t=4$ case this holds \ZamoIII.  Obviously, the
already-ruled-out $\alpha=0$ would work, but there is another interesting
possibility. This is to set $\alpha=\pi/2\g$. We are also free to multiply by
an overall function, so we set
\eqn\slr{S_{RL}(\t)={ \tilde Z(\t)\over Z(\t+{i\pi\over 2\g})}
S_{LL}(\t+{i\pi\over 2\g})
\quad\quad S_{LR}(\t) =-{\tilde Z(\t)\over Z(\t-{i\pi\over 2\g})}
S_{LL}(\t-{i\pi\over 2\g}).}
This replaces $Z$ with $\tilde Z$ and $\sinh$ with $i\cosh$ in each
of \LLRSOS.

\subsec{``Undoing'' the quantum-group truncation}

We could find $Z(\theta)$ and $\tilde Z(\t)$ directly by demanding the usual
constraints of unitarity and crossing symmetry.  However, we first find them
for a related, simpler $S$-matrix.  A well-known ``reduction'' procedure then
gives the RSOS $S$-matrix from this $S$-matrix \refs{\Pasquier,\Leclair}. In
the context of integrable lattice models, this is known as the vertex-IRF
correspondence. The unreduced $S$-matrix we find will also be used in the next
section, where we study the sine-Gordon model when the coefficient of the
potential imaginary. For the moment one can think of it as an intermediate
object that makes the study of normalization convenient.

The unreduced $S$-matrix describes the scattering of two doublets, one
left-moving and the other right-moving.  For the $LL$ or $RR$ scattering, the
only possible $S$-matrix for a doublet $(u,d)$ consistent with factorizability
\YBE\ is the usual sine-Gordon $S$-matrix \ref\ZandZ{A.B.Zamolodchikov,
Al.B.Zamolodchikov, Ann. of Physics 120 (1979) 253.}
\eqn\LLsmat{\eqalign{a&= Z(\theta)\sinh\g(\t-i\pi)\cr
b&= Z(\theta)\sinh\g\t.\cr
c&= -Z(\theta) \sinh i\g\pi\cr}}
where $a$ describes the process $uu\rightarrow uu$, $b$ describes
$ud\rightarrow ud$, $c$ describes the reflection $ud\rightarrow du$ and there
is a symmetry under $u\leftrightarrow d$.  Again, $\g=1/(t-1)$.  Unitarity and
crossing fix $Z$ to be
\eqn\ZLL{Z(\t)= {1\over
 \sinh\g(\theta-i\pi)}\exp {i\over 2}
\int_{-\infty}^{\infty}{dk\over k} \sin k\theta{\sinh {k\pi\over
2}({1\over\g}-1)\over \sinh {k\pi\over 2\g}\cosh {k\pi\over 2}}}
This generalizes to the XXZ case the massless $LL$ $S$-matrix for the XXX
model ($\g\rightarrow 0$) \XXX. It also arose in the WZW models of \ZamZam.

The unreduced $S_{RL}$ then follows from \slr. (This $S$-matrix has the
slightly odd property of having $c_{RL}=-c_{LR}$.)  To fix $\tilde Z(\t)$, we
use the unitarity and crossing relations \refs{\ZamoIII,\ZamZam}. These are
\eqn\unitcross{\eqalign{&a_{LR}(\theta)a_{RL}^*(\theta)=
b_{LR}(\theta)b_{RL}^*(\theta)+c_{LR}(\theta)c_{RL}^*(\theta)=1\cr
&b_{LR}(\theta)c_{RL}^*(\theta)+c_{LR}(\theta)b_{RL}^*(\theta)=0\cr
&b_{RL}(\theta)a_{RL}(i\pi + \theta)+c_{RL}(\theta)c_{RL}(i\pi+\theta)=1\cr
&b_{RL}(\theta)c_{RL}(i\pi + \theta)+c_{RL}(\theta)a_{RL}(i\pi+\theta)=0,\cr}}
where the first two apply only for $\theta$ real, and the last two also hold
true for $LR$-type elements. The second and fourth are satisfied for any
$\alpha$ and $\tilde Z$. The other two give
\eqn\unit{\tilde Z(\theta)\tilde Z^*(\theta)={1\over
\cosh\g(i\pi-\t) \cosh\g(\t+i\pi)}
=\tilde Z(\theta)\tilde Z(i\pi+\theta)}
The simplest solution of these is
\eqn\forZ{\eqalign{\tilde
Z(\theta)=&{1\over \cosh\g(\t-i\pi)}\times\cr
\prod_{n=1}^{\infty}
&{\Gamma^2(\half+(2n-1)\g+{\g\theta\over
i\pi})\Gamma(\half+(2n-2)\g-{\g\theta\over
i\pi})\Gamma(\half+2n\g-{\g\theta\over i\pi})\over
\Gamma^2(\half+(2n-1)\g-{\g\theta\over
i\pi})\Gamma(\half+(2n-2)\g+{\g\theta\over i\pi})
\Gamma(\half+2n\g+{\g\theta\over i\pi})}\cr = & {1\over \cosh\g(i\pi-\t)}
\exp -{i\over 2}\int_{-\infty}^{\infty}{dk\over k}
\sin k\theta{\sinh {k\pi\over 2}\over
\sinh {k\pi\over 2\g}\cosh {k\pi\over 2}}.\cr}}

To find the appropriate RSOS $S$-matrix, we rewrite the $LL$ $S$-matrix as
\eqn\Sinegordon{S_{LL}=Z(\tht)
\left[\sinh\g(\t-i\pi)P
+\sinh(\g\tht)e\right],}
where $P$ is the permutation matrix, and $e$ is
$$\pmatrix{0&0&0&0\cr 0&1&-e^{-i\pi\g}&0\cr 0&-e^{i\pi\g}&1&0\cr
 0&0&0&0\cr}.$$
The $S$-matrix \Sinegordon\ is equivalent to \LLsmat, but we have redefined
the states $|u(\t_i)\rangle\rightarrow e^{-\t_i/2}|u(\t_i)\rangle$ and
$|d(\t_i)\rangle\rightarrow e^{\t_i/2}|d(\t_i)\rangle$.  This ensures that $e$
obeys the Temperley-Lieb algebra \Baxter. Any $S$-matrix of the form
\Sinegordon\ satisfies \YBE\ as long as $e$ obeys this algebra. Our RSOS
$S$-matrix is given by \Sinegordon, but with an $e$ which satisfies the same
algebra but acts on the RSOS particles. The elements of this new $e$ are
$$e={\wgtc abcd} =-\delta_{bd} {\left[\sin\bigl(a\g\pi \bigr)\,
\sin\bigl(c\g\pi \bigr)\right]^{1/2}\over
\sin(b\g\pi)}$$
where adjacent vacua must differ by $\pm 1$. This gives \LLRSOS\ after we do
another rapidity-dependent ``gauge transformation'' to get the $(\beta_m)^{
i\t/\pi}$ and the minus sign. $S_{LR}=S_{RL}$ follows from \slr.

We have completely determined the RSOS $S$-matrix, because we now know $Z(\t)$
from \ZLL\ and $\tilde Z(\t)$ from \forZ.  We can check that it agrees with
the previously-known case $t=4$ ($\g=1/3$) \ZamoIII.  In this case, the RSOS
structure is trivial: all one can do is go back and forth between the two
vacua.  Thus the spectrum consists of one left-moving and one right-moving
particle: the Goldstinos of \ZamoIII.  The two S-matrix elements $S_{LL}$ and
$S_{RL}$ follow from either of the last two relations in \LLRSOS. To show that
this $S$-matrix agrees with that of \ZamoIII, one uses the relation
\eqn\FT{{\sinh\g(a\t+i\pi)\over \sinh\g(a\t-i\pi)}=
\exp -i\int_{-\infty}^{\infty}{dk\over k} \sin k\theta{\sinh {k\pi\over
2a}({1\over\g}-2)
\over\sinh {k\pi\over 2a\g }},}
which is good for $a>0$ and $0\le\g<\half$. This yields $S_{LL}=1$ and
$S_{RL}=-\tanh(\t/2-i\pi/4)$, as required.

To check that these $S$-matrices indeed describe the scattering for all the
flows between minimal models, we will calculate the ground-state energy $E(R)$
on a circle of radius $R$.  At the ultraviolet and infrared fixed points this
is proportional to the central charge $c$ of the appropriate field theory by
using the relation $E(R)=\pi c/6R$. This calculation was done in \ZamoIII\ for
the case $t=4$; in addition, the appropriate equations for all $t$ were
conjectured there. We will derive the equations of \ZamoIII\ explicitly for
$t=5$.

\subsec{TBA}

The technique we will use to calculate the ground-state energies from the
$S$-matrix is known as the thermodynamic Bethe ansatz (TBA) \ZamoI. The TBA
gives the free energy $f=-\ln Z/R$ of a particle gas on a line of large length
$T$ at a temperature $1/R$. This is equivalent to considering a spacetime of a
cylinder of radius $R$ and length $T$. If we change our point-of-view and
think of the spatial direction as (Euclidean) time, only the ground state
contributes to $Z$, since only the lowest-energy state can propagate over the
large distance $T$. Thus $Z=\exp(-TE(R))$.  The equivalence between the two
approaches (often called modular invariance) means that $E(R)=-\ln Z/T=R f/T$.

To calculate the free energy, we need to diagonalize the ``transfer-matrix'',
meaning that we find the eigenvalues for scattering one particle through an
ensemble of all the others. Since the $t=4$ massive case and our $t=5$
massless case are very similar, we first recap the results for the $t=4$
massive case \ZamoII. There are three degenerate wells labeled ($+$, $0$ and
$-$) with a massive kink interpolating between each adjacent pair.  The
$S$-matrix elements are given by \LLRSOS\ with $\g=1/4$. The transfer matrix
$T(\t|\t_1,\t_2,\dots\t_N)$ can be represented pictorially by
\bigskip
\noindent
\hbox{$\hskip2cm$
\rlap{\raise1.0cm\hbox{$\hskip1.15cm 0\hskip1.2cm
\pm\hskip 1.2cm 0\hskip 2.5cm \pm\hskip 1.15cm 0$}}
\rlap{\raise.2cm\hbox{$\hskip1.0cm \pm\hskip1.3cm
0\hskip 1.2cm \pm\hskip 2.5cm 0\hskip 1.2cm \pm$}}
\rlap{\lower10pt\hbox{$\hskip1.5cm \t_1 \hskip1.4cm \t_2\hskip3.8cm
\t_N$}}
{\raise.6cm\hbox{$\t$}}
$\hskip.3cm$\rlap{\raise.75cm\hbox{{\vrule width3.5cm height .4pt}
$\hskip.2cm$ \cc $\hskip.2cm$ \cc $\hskip.2cm$ \cc $\hskip.2cm$
{\vrule width1.9cm height .4pt}}}
$\hskip.75cm$ \hbox{\vrule width.4pt height1.5cm} $\hskip1.35cm$
\hbox{\vrule width.4pt height1.5cm} $\hskip4cm$
\hbox{\vrule width.4pt height1.5cm}}
\bigskip
\noindent
where the intersections represent $S$-matrix elements, and the $\pm$ denote
vacua which can be either $+$ or $-$. We impose periodic boundary conditions,
so $N$ must be even and the rightmost $+$ or $-$ is the same as the leftmost.
This is equvalent to the lattice Ising model transfer matrix: the
configurations are specified by the $\pm$, and the $S$-matrix elements are the
same as Ising-model Boltzmann weights.

To find the transfer matrix eigenvalues $\Lambda(\t|\t_1,\t_2\dots\t_N)$, we
follow \Baxter\ and \ZamoII. We can ignore the $(\beta_m)^{\pm i \t/\pi}$
pieces, because they cancel out in $T(\t|\t_1,\t_2,\dots,\t_N)$.  Using the
weights \LLRSOS\ one finds that if one ignores the $Z(\t)$, shifting
$\t\rightarrow \t+2i\pi$ is equivalent to flipping ($+\leftrightarrow -$) one
of the two Ising spins in a weight and multiplying the weight by $i$. In the
transfer matrix this corresponds to flipping, say, the bottom row of spins.
Thus
\eqn\periodic{\lambda(\t|\{\t_i\})\equiv{\Lambda(\t|\t_1,\t_2\dots\t_N)
\over\prod_{i=1}^N Z(\tht-\tht_i)}=r (-1)^{N/2}\,\l(\t+2i\pi|\{\t_i\}), }
where $r$ is the eigenvalue ($\pm 1$) of the configuration under spin-flip.
We note that $\exp(N\t/4)\l(\t|\{\t_i\})$ is a polynomial of order $N$ in
$\exp(\t/2)$, and by the relation \periodic, the odd terms vanish for
$r=(-1)^{N/2}$ and the even ones for $r=-(-1)^{N/2}$. Thus $\l(\t|\{\t_i\})$
must have $N/2$ zeros in the strip $-\pi\le Im \t\le\pi$ in the first case and
$N/2-1$ in the second. By analyticity, it can thus be written as
\eqn\eval{\lambda(\t|\{\t_i\})=\cases{A\prod_{i=1}^{N/2}
\sinh({\t-x_i\over 2})&
$r=(-1)^{N/2}$\cr A\prod_{i=1}^{N/2-1} \sinh({\t-x_i\over 2})&
$r=-(-1)^{N/2}$\cr}}
where $A$ is independent of $\t$. Specifying the zeros $\{x_i\}$
(which depend on the rapidities $\t_1,\t_2\dots\t_N$) specifies the
eigenvalue. To find the zeros, we use the ``inversion'' relation \foot{This
relation is not hard to derive: one first finds $T(\t)T(\t+i\pi)$ explicitly
for $N=2$ but without periodic boundary conditions. One then builds up the
relation for any $N$ with periodic boundary conditions.} derived in \Baxter:
\eqn\inv{\l(\t|\{\t_i\})\l(\t+i\pi|\{\t_i\})=
\prod_{i=1}^N \cosh({\t-\t_i\over 2}) +
r \prod_{i=1}^N \sinh({\t-\t_i\over 2}).}
Thus if $x$ is a zero it must satisfy $\prod_{i=1}^N \tanh((x-\t_i)/2)=
-r(-1)^{N/2}$.  The $N$ solutions to this polynomial equation in the strip are
of the form $x_i=y_i\pm i\pi/2$, where $y_i$ is real. The $N/2$ (or $N/2-1$)
zeros $\{x_i\}$ needed to find an eigenvalue $\l(\t|\{\t_i\})$ in \eval\ are
then specified by choosing one from each pair. (If we choose both the
left-hand-side of \inv\ will not be zero for some other $x_i$, whereas the
right-hand side still will.)

We have found the eigenvalues of the transfer matrix for the $t=4$ massive
case, and one can then calculate the ground-state energy as in \ZamoII.
Generalizing this to our massless $t=5$ case is only slightly more
complicated. The transfer matrix displayed in the diagram above must carry an
additional label corresponding to $L$ or $R$ on each line. The parameter
$\g=1/4$ in both cases so $S_{LL}=S_{RR}$ is the same, and $S_{RL}$ is given
by the shifted \slr.  Thus, just as we did in the Yang-Baxter case, we can
reduce this to the $LL$-only case (the problem just solved) by pulling out the
$\tilde Z(\t)$ as in \periodic\ and shifting the rapidities of all the
right-movers by $2i\pi$.  This shift affects almost nothing of the above
diagonalization. The eigenvalue $\l(\t|\{\t_i\})$ must have the same number of
zeros in the strip, so \eval\ is exactly the same, and the only effect on
\inv\ is to multiply the right-hand-side by $(-1)^{N_R}$.  The eigenvalue
zeros $x_i=y_i\pm i\pi/2$ are therefore the solutions of
\eqn\fory{\prod_{j=1}^{N_L} {\sinh({y-\t_j\over 2}+{i\pi\over 4})\over
\sinh({y-\t_j\over 2}-{i\pi\over 4})}
\prod_{k=1}^{N_R} {\sinh({y-\t_k\over 2}+{i\pi\over 4})\over
\sinh({y-\t_k\over 2}-{i\pi\over 4})}
=-r(-1)^{(N_L-N_R)/2}}

With this diagonalization complete, we can find relations for the level
densities and particle densities. The total numbers of left and right
particles are each conserved, so we define the densities of states $P_L(\tht)$
and $P_R(\tht)$. In addition, we define the density $P(y)$ of solutions to
\fory. A given particle configuration (eigenstate of the transfer matrix) is
specified by the distributions of rapidites for $L$ and $R$, and the
distributions of eigenvalue zeros for $x_i=y_i+i\pi/2$ and for
$x_i=y_i-i\pi/2$: we thus define the associated densities $\rho_L(\t)$,
$\rho_R(\t)$, $P_{+}(y)$ and $P_-(y)$.  As we have shown,
$P(y)=P_-(y)+P_+(y)$.  Taking the derivative of the logarithm of \fory\ gives
\eqn\Py{2\pi P(y)=\int d\t \ {\rho_L(\t)+\rho_R(\t)\over \cosh (y-\t)}.}
To find $P_L(\t)$ and $P_R(\t)$, we quantize the momenta: the periodic
boundary conditions require that for an eigenstate of the transfer matrix
\eqn\quant{e^{ip_i l}\Lambda(\tht_i| \tht_1, \tht_2,\dots,\tht_N)=1.}
Taking the derivative of the logarithm of \quant\ and using \eval\ gives
\eqn\PL{\eqalign{2\pi P_L(\tht)=
{m\over 2}le^{-\tht}+&{d\over d\tht} Im\ln
\Lambda(\tht|\tht_1,\dots,\tht_N)\cr
={m\over 2}le^{-\t}+&\int d\t^{\p} \phi(\t-\t^{\p})\rho_L(\t^{\p})
+\int d\t^{\p} \tilde\phi(\t-\t^{\p})\rho_R(\t^{\p})\cr
+&\half\int dy {P_+(y)-P_-(y)\over \cosh (y-\t)},\cr}}
where
\eqn\forphi{\eqalign{\phi(\t)=&Im {d\over d\t} \ln Z(\t)={1\over 8}
\int dk \cos k\t {1\over \cosh^2{\pi k\over 2}}\cr
\tilde\phi(\t)=&Im {d\over d\t} \ln \tilde Z(\t)=\phi(\t),}}
where we have used the relation \FT. The explicit expressions for $\phi$ and
$\tilde\phi$ and the relation $\phi=\tilde\phi$ hold only for this $\g=1/4$
case.  Replacing $\exp(-\t)$ with $\exp\tht$ and interchanging $\rho_L$ with
$\rho_R$ gives $P_R$. We get rid of $P_-(y)$ in these expressions by
substituting $P(y)-P_+(y)$ for it; using \Py\ and the relation
$$\int {dy\over 2\pi }{1\over \cosh (y-\t )\cosh y}={1\over 4}\int dk {\cos \t
k \over \cosh ^2({\pi t\over 2})}=2\phi(\t)$$
giving us the remarkably simple relation
\eqn\PLii{2\pi P_L(\tht)={m\over 2}le^{-\tht}+
\int dy {P_+(y)\over \cosh (y-\t)};}
$P_R(\t)$ is the same with $\exp-\t$ replaced with $\exp\t$.

We can now find the free energy and hence the ground-state energy by
minimizing the free energy subject to the constraints \Py\ and \PLii\
\refs{\ZamoI,\ZamoII,\ZamoIII}.  This is standard by now; we provide a few
more details in the appendix. The result is that the ground-state energy is
\eqn\fe{E(R)=- {M\over 4\pi }\int d\t
(e^{\t} \ln (1+e^{-\e _R(\t)}) + e^{-\t} \ln (1+e^{-\e _L(\t)}))}
where the $\e_a$ obey the integral equations
\eqn\TBA{\eqalign{
\e _L(\t )&={MR\over 2} e^{-\t}-\int {d\t ^{\p}\over 2\pi} {1\over
\cosh(\t-\t^{\p})}\ln(1+e^{-\e _1(\t ^{\p})})\cr
\e _1(\t )&=-\int {d\t ^{\p}\over 2\pi} {1\over
\cosh(\t-\t^{\p})}\left(\ln(1+e^{-\e _L(\t ^{\p})})
+\ln(1+e^{-\e _R(\t ^{\p})})\right)\cr
\e_R(\t)&=\e_L(-\t).\cr}}
This set of integral equations is summarized by the diagram
\medskip\noindent
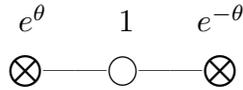
\centerline{\hbox{\rlap{\raise15pt\hbox{$\ \,\,
e^{\theta}\hskip.79cm 1\hskip.69cme^{-\theta}$}}}
$\bigotimes$------$\bigcirc$------$\bigotimes$}
\bigskip
\noindent
This is the result conjectured in \ZamoIII. The properties of these equations
were analyzed in \refs{\ZamoIII,\KM}, so our only comment is that the
effective central charge $c=6RE(R)/\pi$ does indeed start at $4/5$ in the UV
and decreases smoothly to $7/10$ in the IR.

For general $t$, the conjecture of \ZamoIII\ is that the TBA system
\medskip\noindent
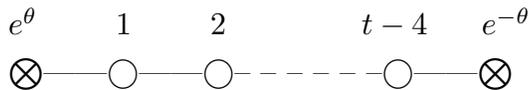
\centerline{\hbox{\rlap{\raise15pt\hbox{$\  e^{\t}\hskip.87cm 1\hskip.87cm
2\hskip1.5cm t-4\hskip.6cm e^{-\t}$}}
$\bigotimes$------$\bigcirc$------$\bigcirc$-- -- --
--  -- --$\bigcirc$------$\bigotimes$}}
\bigskip
\noindent
describes these flows.  This does have all the required properties, so we
expect that doing the TBA calculation with our $S$-matrix will indeed give
this result. Proving this is somewhat tricky. Following \ref\BR{V.V.  Bazhanov
and N.Y.  Reshetikhin, Int. J.  Mod. Phys. A4 (1989) 115.}, one can obtain
equations for the eigenvalues and eigenvalue zeros. The problem is that the
equations for the eigenvalue zeros have more solutions than are needed; some
need to be thrown out. One can either do this by hand (as in \BR) or by first
studying the untruncated $S$-matrix and then obtaining the RSOS result by
adding appropriate chemical potentials to change the boundary conditions
around the circle \FS. For the massive case, this enables the proof of the TBA
equations for all $t$, but in the massless case, various technical
complications arise with either method. It is possible to complete this proof,
but it is complicated and we defer it to the future.  However, in the infrared
limit, $S_{RL}\rightarrow 1$ and the left and right particles decouple. The
TBA calculation then is virtually identical to that of \FS, and we obtain the
correct IR central charge. We will do this calculation in the next section; we
take it as a strong piece of evidence that our $S$-matrices are correct for
all $t$.

\newsec{The sine-Gordon model with imaginary potential}

In the previous paper \FSI, we discussed a number of properties of the
sine-Gordon model
$$S_{\hbox{SG}}
=\int d^2 x\ \left[ \half (\del\phi)^2 + \mu^2 \cos \beta_{SG}
\phi\right]$$
at imaginary $\mu^2$. This describes the continuum limit of the
low-temperature phase of the $O(n)$ lattice model.  It has the free-boson
critical point at $\mu^2=0$, but it has another fixed point which in the
$O(n)$ model corresponds to low temperature. In the phase between the two
critical points, the excitations must be massless; these are the only ones
which remain in the infrared limit.  One way of studying this model is to
analyze the ground-state energy on a cylinder $E(R)$; in \FSI\ we discussed
this by analytically continuing the high-temperature results and by lattice
simulations. In an integrable model such as this, one calculates $E(R)$ and
the effective central charge $c(MR)\equiv 6RE(R)/\pi$ from the $S$-matrix by
the TBA.  This model has central charges of $1$ in both the ultraviolet limit
and in the infrared limit \ref\Isz{C. Itzykson, H.  Saleur and J.B.  Zuber,
Europhys.  Lett. 2 (1986) 91}.

The most plausible guess for the $S$-matrix is given by the untruncated
$S$-matrix in \LLsmat\ and \slr, where $n$ and $\beta_{SG}$ are related to $t$
by
\eqn\param{n=2\cos\left({\pi\over t}\right)=
2\cos{\g\pi\over \g +1};\qquad\qquad \beta_{SG}^2={8\pi t\over t+1},}
and where $M\propto |\mu|^{(t+1)/2}$.  This is for a number of reasons.  The
most compelling is that, as we showed in the previous section, this $S$-matrix
can be reduced to give the minimal-model $S$-matrix; we expect this behavior
from sine-Gordon at imaginary coupling, as was shown from large-$t$
perturbative arguments in \Gri\ (the exact symmetry to be used for this
reduction may however not be the obvious one: this is a delicate issue which
we discuss in sect.\ 4).

Taking some simplifying limits gives further support for this $S$-matrix.
When $\mu\rightarrow 0$ ($t\rightarrow\infty$ and $n=2$), the $RL$ scattering
is trivial. The $L$ and $R$ systems are decoupled for all $M$, not just in the
IR limit.  This system has an ordinary $SU(2)$ symmetry, and was studied in
\ZamZam. This system has $c(MR)=1$ for all masses; this is what we expect from
the $1/t$ expansion done in \FSI. In the other extreme limit $n=-2$
($\g\rightarrow\infty$), the pathologies of the $O(-2)$ model \FSI\ lead one
to expect that the $S$-matrix description breaks down here, as indeed happens
in our $S$-matrix. As a final support for this $S$-matrix we now present
computations at zero temperature but with a magnetic field, which turns out to
be an easier problem.

\subsec{Adding a magnetic field}

\nref\zrefi{G. Japaridze, A. Nersesyan and P. Wiegmann,
Nucl.Phys. B230 [FS10] (1984)
511; P.Wiegmann, Phys. Lett. B152 (1985) 209.}
\nref\zrefii{
P. Hasenfratz, M. Maggiore and F. Niedermayer, Phys. Lett. B245 (1990) 522;
P. Hasenfratz and F. Niedermayer, Phys. Lett. B245 (1990) 529.}
\nref\zrefiii{ Al.Zamolodchikov, ``Mass scale in sine-Gordon and its
reductions'', LPM preprint (1993).}
\nref\zrefiv{M. Ganin, Izv. Vuzov (Math) 33 (1963) 31.}
\nref\zrefv{C. Destri and H. deVega, Nucl. Phys. B358 (1991) 251.}
%\font\ninerm=cmr9
%\font\eightrm=cmr8
%\font\tit=cmr10 scaled \magstep2
%\font\nom=cmr10 scaled \magstephalf

We start with the sine-Gordon action coupled to an external $U(1)$ gauge
field $A_\mu$:
\eqn\zi{S(A)=\int\left[{1\over
2}(\partial_\mu\varphi)^2+\mu^2\cos\beta_{SG}\varphi\right]
d^2x+\int A_\mu j_\mu d^2x .}
Here $j_\mu$ is the $U(1)$ current of the sine-Gordon model
$$
j_\mu={\beta_{SG}\over 2\pi}\epsilon_{\mu\nu}\partial_\nu\varphi
$$
normalized so that in the ordinary ($\mu^2$ real) sine-Gordon model,
\eqn\ziii{
Q=\int j_0 dx_1={\beta_{SG}\over 2\pi}
\int_{-\infty}^\infty{\partial\varphi\over\partial x_1}
dx_1}
is the integer-valued soliton topological charge (the soliton (antisoliton)
has then $Q=1$ ($-1$)). Let us take $A_\mu$ constant (i.e., space-time
independent) and consider the corresponding specific vacuum energy
${\cal E}(A,\mu^2)$ as a function of the ``magnetic'' field $A=|A_\mu|$.

Before we turn to the scattering theory, it is worth looking at the action
\zi\ from the perturbative (in $\mu^2$) point of view. Dimensional arguments
as well as explicit perturbative calculations show that the magnetic field
works as an infrared (IR) cutoff at scales $\sim A$ and therefore if $A\gg
\mu^{1+t}$ the theory is in the ultraviolet regime.  As a leading
$A\to\infty$ approximation we set $\mu=0$ in (1). Then a one-line calculation
gives for the ground-state energy density
\eqn\ziv{{\cal E}(A,0)=-{\beta_{SG}^2\over 8\pi^2}A^2.}
In general it is convenient to define the effective central charge
$k(A,\mu^2)$ as
\eqn\zv{{\cal E}(A,\mu^2)=-{A^2\over \pi}k(A,\mu^2).}
It is plain from the scaling considerations that $k(A,\mu^2)$
is actually a function of the dimensionless variable
$$
\xi=\mu^2/A^{2/(1+t)}
$$
and the perturbative series is in fact an expansion in $\xi^2$
\eqn\zvii{
k_{\rm pert}(\xi)=\sum_{l=0}^\infty k_{2l}\xi^{2l},}
where in particular ($t$ is related to the sine-Gordon parameter $\beta$ as
in \param)
$$
\eqalign{
k_0=&{t\over 1+t}\cr
k_2=&{\pi^2\over 4}\left({2t\over 1+t}\right)^{2(t-1)/(t+1)}
{\displaystyle{\Gamma\left({1-t\over 1+t}\right)}\over
\displaystyle{\Gamma\left({2t\over 1+t}\right)}.}}
$$
At this point let us suppose that the series \zvii\ has some finite radius of
convergence $\xi_0$ defining therefore an analytic function $k_{\rm pert}
(\xi)$ at $|\xi|<\xi_0$. In view of the subsequent Bethe ansatz analysis this
suggestion seems plausible.  We also call ${\cal E}_{\rm
pert}(A,\mu^2)$ the ground state energy defined by this perturbative series
\eqn\zix{{\cal E}_{\rm pert}(A,\mu^2)=-{A^2\over\pi}k_{\rm pert}(\xi).}
Note that this perturbative definition does not care if $\mu^2$ is real or
imaginary; it is independent of the physics of the corresponding field
theories.

We now discuss the scattering theory. Let us choose $x_1$ as the space-like 2D
coordinate and direct the constant field $A_\mu$ along the ``imaginary-time''
coordinate $x_0$. The magnetic field perturbation in action \zi\ modifies
the sine-Gordon Hamiltonian ${\cal H}_{\rm sG}$ to be
\eqn\zx{{\cal
H}(A)={\cal H}_{\rm sG}-AQ}

Consider first the unitary massive sine-Gordon model ($\mu^2$ real in \zi) and
let $m$ be the mass of the corresponding charged particle (soliton).  As usual
the on-mass-shell momenta $(e,p)$ are parameterized in terms of rapidity
$\theta$
$$ e=m\cosh\theta\ ;\ \ \ \ \ \ p=m\sinh\theta $$
Under \zx\ every soliton (antisoliton) acquires additional energy $A$ ($-A$).
It is clear that if $A>m$ the state without particles is no longer the
ground state. The true vacuum contains a sea of positively-charged solitons
which fill all possible states inside some ($A$-dependent) ``Fermi interval''
$-B<\theta<B$.  There is non-trivial scattering between the solitons which
certainly influences the structure of the ground-state sea.  However, only one
kind of particle participates and we can handle their factorized scattering as
a diagonal one, the two-particle amplitude being $-a(\theta)$ of \LLsmat\ with
$$\mu=1/t$$
for the moment. Along the standard lines of the Bethe ansatz (BA) technique
(see e.g.\ \refs{\zrefi-\zrefiii}) one arrives at the following expression for
the specific ground state energy ${\cal E}^{(Re)}(A)$ (the superscript $Re$ is
added to stress that currently we address the ordinary sine-Gordon with real
coupling $\mu^2$)
\eqn\zxiii{{\cal E}^{(Re)}(A)-{\cal E}^{(Re)}(0)= -{m\over
2\pi}\int_{-B}^B\cosh\theta\ \rho(\theta)d\theta.}
The positive function $\rho(\theta)$ (the particle density in the ground
state) is defined inside the Fermi interval $-B<\theta<B$ and solves the BA
equation
\eqn\zxiv{A-m\cosh\theta=\rho(\theta)-\int_{-B}^B
\phi(\theta-\theta')\rho(\theta')d\theta'}
while the Fermi boundary $B$ is determined by the boundary condition
\eqn\zxv{\rho(\pm B)=0.}
The kernel $\phi(\theta)$ follows from the soliton-soliton
scattering:
\eqn\zxvi{\phi(\theta)={1\over 2\pi i}{d\over d\theta}\log a(\theta)=
\int{e^{i\omega\theta}\sinh{\pi(t-1)\omega\over 2}\over
2\cosh{\pi\omega\over 2}\sinh{\pi t\omega\over 2}}{d\omega\over 2\pi}}

We are primarily interested in the UV ($A\to\infty$) analytic structure of
${\cal E}^{(Re)}(A)$. Since $B\to\infty$ in this limit, it is convenient to
transform (14) to the following Wiener-Hopf equation
\refs{\zrefiv,\zrefi-\zrefiii}
\eqn\zxvii{
v(\omega)=-{iAK_+(0)\over\omega}+{ime^B\over 2}{K_+(-i)\over\omega-i}+
\int_{C_+}{e^{2i\omega' B}\over\omega+\omega'}\alpha(\omega')v(\omega')
{d\omega'\over 2\pi i}}
for the meromorphic function $v(\omega)$. Here
\eqn\zxviii{\eqalign{
K_+(0)&=\sqrt{2t/(1+t)}\cr
K_+(-i)&=\sqrt{2\pi\over t}\left({t\over 1+t}\right)^{t/2}{\displaystyle{
\Gamma\left({1+t\over 2}\right)}\over\displaystyle{\Gamma\left({t\over 2}
\right)}}}}
and the kernel function $\alpha(\omega)$ reads explicitly
\eqn\zxix{\alpha(\omega)=e^{2i\omega\Delta}
{\displaystyle{\Gamma\left(-{it\omega\over 2}\right)\Gamma\left({1\over 2}-
{i\omega\over 2}\right)\Gamma\left({i(1+t)\omega\over 2}\right)}\over
\displaystyle{\Gamma\left({it\omega\over 2}\right)\Gamma\left({1\over 2}+
{i\omega\over 2}\right)\Gamma\left(-{i(1+t)\omega\over 2}\right)}}}
with
$$
\Delta={t\over 2}\log t-{1+t\over 2}\log(1+t).
$$
The integration contour $C_+$ encircles all singularities on the positive
imaginary axis. These are only the poles of $\alpha(\omega)$ at
\eqn\zxxi{\omega={2in\over 1+t}\ ;\ \ \ \ \ \ n=0,1,2,\ldots}
The boundary condition \zxv\ now reads as
\eqn\zxxii{
iAK_+(0)-{ime^B\over 2}K_+(-i)=\int_{C_+}e^{2i\omega B}\alpha(\omega)
v(\omega){d\omega\over 2\pi  i}}
while the ground state energy \zxiii\ becomes
\eqn\zxxiii{\eqalign{
{\cal E}^{(Re)}(A)-{\cal E}^{(Re)}(0)&=
-{me^B\over 2\pi}K_+(-i)\ \times\cr
&\left[AK_+(0)-{me^B\over 4}K_+(-i)+\int_{C_+}{e^{2i\omega B}\over
\omega-i}\alpha(\omega)v(\omega){d\omega\over 2\pi i}\right]\cr}}
In the last equation the integration contour contains one extra pole at
$\omega=i$ which one can pick up explicitly. We recognize its contribution
${1\over 4}m^2\tan(\pi t/2)$ as (with the opposite sign) the well-known bulk
vacuum energy (without any magnetic field) of the sine-Gordon model
\refs{\zrefiii,\zrefv}, i.e.\ what we have denoted ${\cal E}^{(Re)}(0)$ in
eqn.\ \zxiii
\eqn\zxxiv{
{\cal E}^{(Re)}(0)=-{m^2\over 4}\tan{\pi t\over 2}.}
The bulk term looks infinite for $t$ an odd integer because in this case it
depends on $m^2 \ln m$ \ZamoIII.
The remaining part of \zxxiii, using the scaled definition \zv,
develops as a regular expansion in powers of $(m/A)^{4/(1+t)}$ i.e.\
\eqn\zxxvi{k^{(Re)}(A,m)=\sum_{l=0}^\infty K_l\ (m/A)^{4l/(1+t)}}
This series should be the same as the perturbative one \zvii\ provided
that \zrefiii
\eqn\zxxvii{{\pi|\mu^2|\over 2}=
{\displaystyle{\Gamma\left({t\over 1+t}\right)}\over
\displaystyle{\Gamma\left({1\over 1+t}\right)}}
\left[{m\sqrt{\pi}\over 2}
{\displaystyle{\Gamma\left({1+t\over 2}\right)}\over
\displaystyle{\Gamma\left({t\over 2}\right)}}\right]^{2/(1+t)}}

{}From a formal point of view one may interpret the BA system \zxiii--\zxv\ as
a convenient way of summing up the perturbative expansion \zvii\ at real
$\xi$. One would like a similar tool to sum up \zvii\ at $\xi$ purely
imaginary, i.e.\ the alternating series. We show here that the massless
scattering theory proposed above provides the necessary information to resum
this series.  Let us turn on the magnetic field perturbation \zx\ (say with
$A>0$) in our new world of massless charged particles. The energy spectrum
$$
\eqalign{
e=p&={M\over 2}e^\theta\ \ \ \ \ \ \ \ \ \ \ \hbox{for right-movers}\cr
e=-p&={M\over 2}e^{-\theta}\ \ \ \ \ \ \ \ \ \hbox{for left-movers}}
$$
is gapless and the positively charged particles (say the $u$ ones) are
always excited in the ground state once $A>0$. Now the right- and
left-movers fill respectively the semi-infinite Fermi intervals
$-\infty<\theta<B$ and $-B<\theta<\infty$ with some Fermi boundary $B\sim
\log A/M$. Again it is a straightforward Bethe ansatz exercise to derive the
following system of integral equations
\eqn\zxxix{\eqalign{
QA-{Me^\theta\over 2}&=
\rho_R(\theta)-\int\limits_{-\infty}^B \phi_1(\theta-\theta')
\rho_R(\theta')d\theta'
-\int\limits_{-B}^\infty \phi_2(\theta-\theta')\rho_L(\theta')
d\theta';\cr
QA-{Me^{-\theta}\over 2}&=
\rho_L(\theta)-\int\limits_{-B}^\infty \phi_1(\theta-\theta')
\rho_L(\theta')d\theta'
-\int\limits_{-\infty}^B \phi_2(\theta-\theta')\rho_R(\theta')
d\theta'.}}
The positive functions $\rho_R(\theta)$ and $\rho_L(\theta)$ are defined in
the Fermi intervals $\infty<\t<B$ and $-B<\t<\infty$ respectively, and are
restricted by the boundary conditions
\eqn\zxxx{\rho_R(B)=\rho_L(-B)=0}
In eqn.\ \zxxix\ we introduced the $U(1)$ charges $\pm Q$ of the massless $u$
and $d$ particles, which we see no reason to fix in advance. The kernels
$\phi_1(\theta)$ and $\phi_2(\theta)$ are related to the LL and RL diagonal
$uu\to uu$ scattering amplitudes respectively (see \LLsmat\ and \slr\ where
now $\mu=1/(t-1)$)
\eqn\zxxxi{\eqalign{
\phi_1(\theta)&={1\over 2\pi i}{d\over d\theta}\log a_{LL}(\theta)=
\int{e^{i\omega\theta}\sinh{\pi(t-2)\omega\over 2}\over
2\cosh{\pi\omega\over 2}\sinh{\pi(t-1)\omega\over 2}}{d\omega\over 2\pi}\cr
\phi_2(\theta)&={1\over 2\pi i}{d\over d\theta}\log a_{RL}(\theta)=
-\int{e^{i\omega\theta}\sinh{\pi\omega\over 2}\over
2\cosh{\pi\omega\over 2}\sinh{\pi(t-1)\omega\over 2}}{d\omega\over 2\pi}.\cr}}
Finally, the ground state energy, which we now call ${\cal E}^{(Im)}(A)$, is
evaluated as follows
\eqn\zxxxii{
{\cal E}^{(Im)}(A)-{\cal E}^{(Im)}(0)=
-{M\over 2\pi}\int_{-\infty}^Be^\theta\ \rho_R(\theta)d\theta,}
where we have taken into account the obvious symmetry $\rho_R(\theta)=
\rho_L(-\theta)$.

Although this new BA system \zxxix--\zxxxii\ has a rather different form from
that of eqns.\ \zxiii--\zxxvi, it is easy to relate the two in the UV region
$A\to\infty$ where in both systems $B\to\infty$. In this limit the right and
left Fermi intervals have a broad overlap at $-B<\theta<B$. Near say the right
Fermi boundary $\theta\sim B$ (where the main contribution to \zxxxii\ comes
from) we can forget about the left one and solve \zxxix\ for $\rho_L(\theta)$
by the Fourier transform with $B\to\infty$. The resulting equation for
$\rho_R(\theta)$ is
\eqn\zxxxiii{rQA-{Me^\theta\over 2}=
\rho_R(\theta)-\int_{-\infty}^B \Phi(\theta-\theta')\rho_R(\theta')
d\theta'}
where in terms of the Fourier transforms
\eqn\zxxxiv{
\tilde \Phi(\omega)=\tilde \phi_1(\omega) +
{[\tilde\phi_2(\omega)]^2\over 1-\tilde
\phi_1(\omega)}= {\sinh{\pi(t-1)\omega\over 2}\over
2\cosh{\pi\omega\over 2}\sinh{t\omega\over 2}}}
(compare with eqn.\ \zxvi) and
$$
r=1+{\tilde \phi_2(0)\over 1-\tilde \phi_1(0)}={t-1\over t}
$$
It coincides precisely with the corresponding limit $B\to\infty$ of eq.(14)
provided
\eqn\zxxxvii{
\eqalign{&Q={t\over t-1}\cr
&M=m\cr}}

For more detailed information we again transform the system \zxxix--\zxxxii\
into Wiener-Hopf form:
\eqn\zxxxviii{
u(\omega)=-rQ{iAK_+(0)\over\omega}+{iMe^B\over 2}{K_+(-i)\over\omega-i}+
\int_{C_+}{e^{2i\omega'
B}\over\omega+\omega'}\overline\alpha(\omega')u(\omega')
{d\omega'\over 2\pi i}}
with the same $K_+(0)$ and $K_+(-i)$ as in eqn.\ \zxviii\ and the same
integration contour $C_+$. Likewise the boundary condition \zxxx\ turns to
\eqn\zxxxix{
irQAK_+(0)-{iMe^B\over 2}K_+(-i)
=\int_{C_+}e^{2i\omega B}\overline\alpha(\omega)
u(\omega){d\omega\over 2\pi i}}
while the ground state energy reads
\eqn\zxl{\eqalign{
{\cal E}^{(Im)}(A)-&{\cal E}^{(Im)}(0)=-{Me^B\over 2\pi}K_+(-i)\ \times\cr
&\left[rQAK_+(0)-{Me^B\over 4}K_+(-i)+\int_{C_+}{e^{2i\omega B}\over
\omega-i}\overline\alpha(\omega)u(\omega){d\omega\over 2\pi i}\right]
}}
With the relations \zxxxvii\ the only difference between the two systems
\zxvii--\zxxiii\ and \zxxxviii--\zxl\ is in the kernel function
which is now a bit modified
\eqn\zxli{\overline\alpha(\omega)
=-{\sinh{\pi\omega\over 2}\over\sinh{\pi t\omega
\over 2}}\alpha(\omega)}
where $\alpha(\omega)$ is that of eqn.\ \zxix. Picking up the pole
$\omega=i$ in \zxl\ we find the bulk vacuum energy of the sine-Gordon model at
imaginary $\mu^2$:
$$ {\cal E}^{(Im)}(0)={M^2\over 4\cos(\pi t/2)} $$
At $t$ integer this is the same bulk term as in
the corresponding (reduced) interpolating flows between the minimal models
\ZamoIII\ (as speculated in \zrefi), a result we used in our
first paper.

The analytic structure of the remaining part of eqn.\ \zxl, i.e.\ of $$ {\cal
E}^{(Im)}(A)=-{(rQA)^2\over\pi}k^{(Im)}(A) $$ is quite regular and controlled
by the poles of the kernel function $\overline\alpha(\omega)$. These form just
the same set \zxxi\ as that of $\alpha(\omega)$ (note that the poles of the
extra multiplier in \zxli\ are all cancelled by the appropriate zeros in
\zxix). Moreover, the residues are almost the same, only the sign alternating
at $n$ odd in \zxxii. This obviously results in the following analytic
structure
$$ k^{(Im)}(A,M)=\sum_{l=0}^\infty
(-)^lK_l\ \left({M\over rQA}\right)^{4l/(1+t)}
$$
with precisely the same coefficients $K_l$ as in expansion \zxxvi.

We conclude that up to the known bulk vacuum energy contributions the massless
BA system \zxxix--\zxxxii\ gives the correct analytic continuation of the
massive one \zxiii--\zxvi\ to purely imaginary $\xi$, providing \zxxxvii\
holds.  In particular, the low-temperature mass scale $M$ is in the same
relation
\zxxvii\ with the absolute value of $\mu^2$.

The success of this computation justifies the $t\rightarrow t-1$ shift, the LR
doubling and the shift of $\theta$ in the LR matrix elements. It seems to make
the above $S$-matrix the right choice for the non-unitary sine-Gordon model.
However, as we shall now see, the TBA does not lead to the correct form of the
ground state energy on a cylinder (i.e.\ the finite-temperature free energy).

\subsec{The case $t=2$}

The $S$-matrix appears natural in the study of the $n$=0 point too. At the UV
fixed point, this model has $N$=2 supersymmetry, but in the low-temperature
phase, the supersymmetry is spontaneously broken (this is possible thanks to
the non-unitarity).  The spectrum should consist of four massless Goldstinos,
one for each supersymmetry. The left movers and the right movers each form a
$(u,d)$ doublet.  At the zero-temperature fixed-point, this model has $c=-2$,
so in the infrared limit, the Goldstinos become the well-known $\eta \xi$
ghost system.  This is a non-unitary version of what happens in the flow
between the tricritical Ising model and the Ising model \ref\KMS{D. Kastor, E.
Martinec and S. Shenker, Nucl. Phys. B316 (1989) 590.} (the $t=4$ case in the
previous section), where a Goldstino resulting from the spontaneously-broken
$N$=1 supersymmetry becomes the Ising-model free-fermion in the infrared
limit. One can write down a Landau-Ginzburg action for the $N$=2 case.  The
$N$=2 superfield has two fermions and two bosons; to obtain the Goldstino
interactions one can integrate out the bosons, as was done in the $N$=1 case
in \KMS. One also expects the $LL$ and $RR$ $S$-matrices to be the identity,
because the $LL$ scattering essentially describes only IR properties, and as
we mentioned above, the $n=0$ system in this limit reduces to the free
$\eta\xi$ system.  $S_{RL}$ diagonal then follows from factorizability. Our
proposed $S$-matrix has these properties for $\mu=1$.

We can now consider the scattering theory at finite temperature and, via the
TBA, extract the ground state energy on a cylinder. Surprisingly, it does not
have the expected behavior. We provide the TBA equations for this diagonal
scattering in the appendix. Although $c_{\hbox{eff}}=1$ in the IR, it is equal
to $.6$ in the UV.

Using this $S$-matrix, we have also calculated $\tr (-1)^F$ in the $t=2$ case.
This corresponds to changing the boundary conditions around the cylinder from
the Neveu-Schwarz sector to Ramond. In a supersymmetric theory, $\ln\tr
(-1)^F$ does not depend on $L$, so it should vanish in a TBA calculation.  We
do this calculation by inserting chemical potentials of $i\pi/R$ for the
Goldstinos, and in the infrared limit it does not vanish but instead gives
$c=-2$. This is not the ground-state energy, but in fact is the actual central
charge of the IR fixed point. (The effective $c=1$ results from subtracting
$1/12$ times the dimension of the lowest-dimension operator in this
non-unitary theory.) When we move away from the IR limit, we find striking
behavior: this quantity has a discontinuity at $mR\approx 3.6$ (near to $r_2$
of the previous paper), where it takes value $c\approx -4.10$. For $mR$
smaller the TBA does not seem to make any sense except when postulating that
there are no particles left, which corresponds to an effective central charge
equal to zero. This strange behavior is actually identical to what was
observed in lattice simulations, where the singularity occurs due to the level
crossing.  Compare in particular with figure 5 of the preceding paper. Thus it
seems that the $S$-matrix leads to the correct ground state energy in the R
sector, but not in the NS sector.

\subsec{General $t$}

To find the ground-state energy using the TBA for general $t$, we must
diagonalize the ``transfer matrix'' as before. This time, we use the algebraic
Bethe ansatz \ref\FST{E.K.  Sklyanin, L.A.  Takhtadzhyan and L.D. Faddeev,
Theo. Math.  Phys. 40 (1980) 688. L. Faddeev, in Les Houches 1982 {\it Recent
advances in field theory and statistical mechanics} (North-Holland 1984),
edited by J.-B.  Zuber and R.  Stora.}. Just as we did in the minimal-model
case, we can reduce this to an all-$LL$ problem by shifting the rapidities of
all the right-movers by $i\pi/2\g$. This is the case studied in \FST, and the
eigenvalues $\Lambda_L(\theta,\{\t_i\})$ and $\Lambda_R(\theta,\{\t_i\})$ are
\eqn\Tmatevs{\eqalign{\Lambda_L (\t ;\{ \theta\},\{ y\})=&
\prod _{j=1}^{N_L}a(\t -\t^L _j)\prod _{k=1}^{N_R}a_{RL}(\t -\t^R _k)
\prod_{r=1}^m {a(y_r-\t)\over b(y_r-\t)}\cr
\Lambda_R (\t ;\{ \theta\},\{ y\})=&
\prod _{j=1}^{N_L}a_{RL}(\t -\t^L _j)\prod _{k=1}^{N_R}a(\t -\t^R _k)
n\prod_{r=1}^m {a_{RL}(y_r-\t)\over b_{RL}(y_r-\t)}\cr}}
where $m$ is an integer $0\le m\le N_L+N_R$, and the $\{y_r\}$ are solutions
of
\eqn\yc{\prod _{j=1}^{N_L}{b(y_r -\t^L _j)\over a(y_r -\t^L _j)}
\prod _{k=1}^{N_R}{b_{RL}(y_r -\t^R _k)\over a_{RL}(y_r -\t^R _k)}=
-\prod_{i=1}^m {a(y_r-y_i)b(y_i-y_r)\over a(y_i-y_r)b(y_r-y_i)}}
One sees that these equations are invariant under exchange of left and right
by making the substitution $y_r\rightarrow y_r+ i\pi/(2\mu)$.

As in the last section, one defines particle densities $\rho_L(\theta)$ and
$\rho_R(\theta)$ for the left- and right-movers. For general $\g$, there are
many more types of solutions of \yc\ than the simple ones for its analog
\fory; these are called string solutions \rTS. One then defines a density for
each type of string solution and then proceeds as before.  There is, however,
a major complication in our case which makes the TBA somewhat more intricate.
We discuss this in the appendix. We were only able to find a result for the
$t=3$ case, and while it has $c_{\hbox{eff}}=1$ in the IR, it is $.2$ in the
UV. Thus this seems to be unsuccessful like the $t=2$ case.

We can, however, analyze these equations successfully in the IR
($M\rightarrow\infty$) limit for $1/\mu$ any integer $\ge 2$, and derive
that $c_{IR}=1$ for the untruncated case, and $c_{IR}=1-6/t(t-1)$ for the RSOS
models.  In this limit, $S_{RL}=1$ and the left and right sectors decouple.
The TBA for each of these sectors is essentially the same as that for the
massive sine-Gordon models discussed in \FS, and is described by the diagram
\bigskip
\noindent
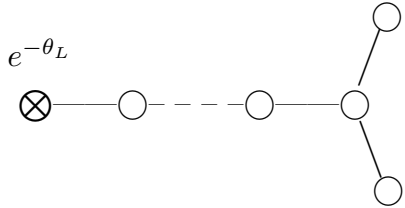
\centerline{\hbox{\rlap{\raise28pt\hbox{$\hskip4.5cm\bigcirc\hskip.25cm $}}
\rlap{\lower27pt\hbox{$\hskip4.4cm\bigcirc\hskip.3cm $}}
\rlap{\raise15pt\hbox{$\hskip4.1cm\Big/$}}
\rlap{\lower14pt\hbox{$\hskip4.0cm\Big\backslash$}}
\rlap{\raise15pt\hbox{$e^{-\t_L}$}}
$\bigotimes$------$\bigcirc$-- -- --
--$\bigcirc$------$\bigcirc$ }}

\bigskip
\noindent for the left, with the same for the right. There are $1/\mu$ open
dots in the diagram. This system has $c_{IR}=1$.

This result allows us to derive $c_{IR}$ for the RSOS $S$-matrices discussed
in the previous section. These RSOS $S$-matrices were obtained from the
untruncated one, and the TBA calculations are identical except that
appropriate twisted boundary conditions must be placed around the cylinder to
reproduce the RSOS model (this is the analog of the charge at infinity in
conformal models). As discussed in \FS, this has the effect of removing the
three nodes at the right end of the diagram. Thus the above diagram becomes
the one at the end of the last section, but with the right-moving particle
removed. This is of course the conjectured answer for the flows between
minimal models, and is a good consistency check on the $S$-matrices discussed
in the last section.

\newsec{Discussion}

One thing we see is clear. At any coupling in sine-Gordon we have $c=1$ in the
IR limit, but at the values $t=2,3$ where we can do the TBA calculation, $c\ne
1$ in the UV (for the trivial case $t\rightarrow\infty$ we still get
$c_{UV}=1$). Thus, at least for $t=2,3$ the $S$-matrix cannot be used to
describe, via the TBA, the flow of the running central charge between the
sine-Gordon fixed points, which have $c=1$ in the IR and UV.

A first possibility is that the $S$-matrix is incorrect for every value of
$t$. However we know that it gives the right TBA after truncation. This seems
compatible only if additional massive particles are present, but disappear in
the truncation.  For this to happen, the particles should have a non-vanishing
charge, and thus affect the magnetic computation. Since we have found this
perfectly consistent, the existence of additional massive particles is
doubtful. Similarly, the computations in the magnetic and truncated cases seem
to exclude the presence of additional CDD factors.

A second possibility is that the $S$-matrix is incorrect only for $t\leq 3$,
which is unfortunately the only range we were able to study with the TBA.  One
hint towards this possibility is that for $t<3$, there is a pole in the
$S$-matrix which in the massive case would correspond to another state in the
spectrum.  For example, for $t=2$, where $S_{RL}=\pm\tanh(\t/2 +i\pi/4)$, this
pole is at $\t=i\pi/2$.  In this massless case it is conceivable that it
corresponds to the massive sine-Gordon particle, which now is an unstable
resonance.  Thus there may be some sort of phase transition at $t=3$, where
this pole crosses the ``physical'' threshold of $\theta=i\pi$.

A variant would be that the $S$-matrix is correct for $t\leq 3$ but cannot be
used to get the running central charge via the TBA. This is conceivable
because the $S$-matrix is defined only in an infinite space-time volume
whereas the TBA applies at finite Euclidan time (except in the IR limit, where
it gives the expected answer). Let us discuss this further.  Consider first
$t=2$.  Recall that $N$=2 supersymmetry is not explicitly broken in the flow.
One therefore could expect that it is realized nonlinearly in the scattering
theory.  However, the fermion number is unambiguously defined along the entire
flow, including at the IR fixed point, and is proportional to to the
topological charge (this can be checked explicitly for various $L$-leg polymer
observables in the dilute and dense cases).  For sine-Gordon solitons in the
massive flow one has $F=\pm 1/2$. For $\eta,\xi$ and their complex conjugates
in the dense $N$=2 case one has $F=\pm 1$.  The value of $F$ for the massless
particles in the $S$-matrix is therefore determined, and $F$ should act
linearly on multiparticle states.  Therefore one has unambiguously that the
Witten index $\tr (-1)^F$ (which gives Ramond boundary conditions) is
expressed in the scattering theory for $t=2$ by adding fugacities $\lambda=-1$
for the fermions. But it is easy to check that with such boundary conditions
in space and time directions, there is no cancellation of various
contributions in the scattering theory, and for any value of $MR$ larger than
$r_2$ the corresponding partition function is not the expected constant, as
discussed at the end of sect.\ 3.2.  This suggests that $N=2$ supersymmetry
cannot be realized at all in our scattering theory.  Such a possibility may
not be surprising.  Recall that even though levels cross the ground state of
the Ramond sector and spontaneously break the $N=2$ supersymmetry, the Witten
index remains equal to 2 along the flow. We set it by hand to zero at the IR
fixed point since the bulk term is there infinite and the levels contributing
to this index are infinitely far from the ground state. Reaching this IR fixed
point is thus a rather singular process, and since the scattering theory is a
perturbation of this IR fixed point, it appears reasonable that it cannot be
used ``to go all the way'' to the ultraviolet model. More precisely we gave in
the previous paper numerical indications that in the Neveu-Schwarz sector the
ground-state energy has singularities at $r_2$ and $r_4$, and maybe at all
values of $r_{2n}$ as well.  If this is true, one may be in a phase that
cannot be described by the TBA and scattering theory. Such a situation is
expected when the virial expansion does not converge \ref\Dashen{R. Dashen,
S.K. Ma, H.J.  Bernstein, Phys. Rev. 187 (1969) 345.}.

 The case $t=3$ is a little marginal.  The associated minimal model is in fact
massive --- it is the Ising model flowing from $c=1/2$ to the trivial $c=0$
fixed point. One does does not expect to find an $S$-matrix describing a
perturbation of the IR fixed point to be able to describe the flow, because it
has only trivial degrees of freedom on which to act. This is reproduced by the
reduction of our sine-Gordon $S$-matrix, which gives a model with no particles
and thus a ground state energy always equal to zero (and not reproducing
$c_{UV}=1/2$).  As an aside notice that $S$-matrices with $U_{q'}sl(2)$
quantum group symmetries acting on spin $1/2$ representations can be
interpreted directly, using another representation of the Temperley-Lieb
algebra, as $S$-matrices describing scattering in an $O(n')$ symmetric theory
with $n'$ analytically continued to $n'=-q'-q'^{-1}$ (for the massive case,
see \refs{\Zamopol,\Smir}). For $t=3$ we have $q'=i$, which gives $n'=0$.  The
result fits with a naive Goldstone picture, where in the low-temperature phase
of the $O(n=1)$ model only degrees of freedom with $O(n'=n-1=0)$ symmetry are
present.

At the present time we can only hope that our $S$-matrix gives the correct
finite-temperature results for $t>3$.

To conclude we discuss briefly the strange symmetry pattern of the problem,
which may be the origin of this confusion.  We expect the minimal model to be
obtained by twisting and reduction of the full sine-Gordon theory, using the
$U_q\widehat{sl(2)}$ quantum symmetry with $q=-\exp (-i\pi/t)$ (the subalgebra
$U_qsl(2)$ with charges of zero dimension in the twisted theory appears on the
lattice). For the scattering theory, this symmetry is expected to act
nonlinearly. On the other hand, the minimal $S$-matrices we have conjectured
above are obviously related to $U_{q'}sl(2)$ symmetry, where
$q'=-\hbox{exp}(-i\pi/(t-1))$, acting linearly.  ``Undoing'' the truncation as
we did is equivalent to assuming that the linear $U_{q'}sl(2)$ symmetry is a
nonlinear realization of the subalgebra $U_qsl(2)$.  In that case the
scattering theory for the full sine-Gordon is indeed obtained by going from
RSOS to vertex basis. This certainly makes sense from the point of view of the
topological charge, i.e.\ the Cartan subalgebra.  Indeed, we showed above that
at the IR fixed point, the central charge of the minimal model was obtained
from the sine-Gordon TBA by introducing fugacities $\lambda=-q'^{\pm 1}$. On
the other hand, in the lattice model as well as for sine-Gordon in the UV, the
central charge of minimal models is obtained by taking traces of evolution
operators with the insertion of $q^{Q}$ where $Q$ is the topological charge
\ziv
$${\beta_{SG}\over 2\pi}\int_{-\infty}^{\infty}dx\ \partial_x\phi.$$
This is of course the soliton number in sine-Gordon. It is as well
proportional to the magnetic number in the Coulomb gas and solid-on-solid
models, the spin in the $O(n)$ vertex model, and the Cartan generator $H(S^z)$
of $U_qsl(2)$. From this one expects
\eqn\corr{Q=\pm {t\over (t-1)}}
for the topological charge carried by the massless particles of our scattering
theory. We found this result independently in the magnetic calculation, so
everything seems consistent. The nonlinear realization of the other generators
remains however to be found. It may be that the mechanism of truncation in
this massless case proceeds very differently from our naive approach. In such
a case, the way we have ``undone'' the quantum-group truncation may be
incorrect, and the non-unitary sine-Gordon $S$-matrix very different.

\bigskip\bigskip\centerline{{\bf Acknowledgments}}\nobreak
We would like to thank K.\ Intriligator, A.\ LeClair, G.\ Moore, C.\ Vafa and
A.B.\ Zamolodchikov for interesting and useful conversations.  H.S.\ was
supported by the Packard Foundation and DOE grant DE-AC-76ERO3075, while P.F.\
was supported by DOE grant DEAC02-89ER-40509.

\appendix{A}{The TBA equations}

Starting with equations for the level densities of the form
\eqn\forrho{2\pi P_a(\tht)= l\nu_a(\t) +
\sum_b \int d\tht'\rho_b(\tht')\phi_{ab}(\theta-\theta'),}
one then minimizes the free energy, and obtains the ground-state energy
\eqn\fe{E_{\l}(R)=-\sum _a {1\over 2\pi }\int d\t\,
\nu_a(\t) \ln (1+\l_a e^{-\e _a(\t)}),}
where the functions $\e_a$ obey the TBA equations \ZamoI
\eqn\TBA{\e _a(\t )=R\nu_a(\t )-\sum _b \int {d\t ^{\p}\over 2\pi}\phi^T
_{ab}(\t -\t ^{\p})\ln(1+\l_b e^{-\e _b(\t ^{\p})}),}
and where the $\l_a$ are fugacities. It is important to notice that the kernel
which appears in \TBA\ is the transpose of that which appears in \forrho;
i.e.\ $\phi^T_{ij}=\phi_{ji}$. In most cases, the kernel is symmetric, but we
will see below that in our sine-Gordon case it is not.  To find the central
charge in the UV limit, one can rewrite \fe\ in terms of a sum of dilogarithms
\ZamoI; this argument only works when $\phi=\phi^T$.

In the simple example $\g=1$ from section 2, the scattering is diagonal so
$\phi_{ab}=-i\del \ln\,S_{ab}(\t)/\del\t$. This leaves us
$\phi_{RL}=-1/\cosh\t$, and $\phi_{LL}=\phi_{RR}=0$. There are two
right-moving particles with $\nu_R(\t)=(M/2) \exp\t$, and two left-movers with
$\nu_L(\t)=(M/2) \exp(-\t)$.  The analysis of dilogarithms then gives
$c_{UV}=.6$ and $c_{IR}=1$. To calculate $\tr(-1)^F$, one includes fugacities
of $-1$ for all the particles.  In the IR limit, this gives $c=-2$; the UV
limit is ambiguous because of the discontinuity of the solution at $mR\approx
3.6$.

We would now like to study the TBA equations for general $\g$. There is a
unusual complication here. Let us study this in the simplest example
$\mu=1/2$, where the left-hand-side of \yc\ is just $(-1)^{N_L+N_R}=1$.
\foot{See \ref\FI{P. Fendley and K.  Intriligator, Nucl.  Phys.  B372 (1992)
533.}, sec 6.1, for a similar calculation done in detail.} Then for large
$N_L+N_R$, the solutions of \yc\ are of the form $y=z_0 +i\pi/2$, or
$y=z_{\bar 0}-i\pi/2$, where $z_0$ and $z_{\bar 0}$ are real. The derivative
of the log of \yc\ then yields
\eqn\Ponetwo{\pm 2\pi P_i(z_i)=\int d\t {\rho_L (\t )\over \cosh (z_i-\t)}-
\int d\t{\rho_R (\t )\over \cosh (z_i-\t)},}
where $i=0$ or $\bar 0$. The $\pm$ in front is normally fixed by demanding
that $P_i$ be positive. This is where the complication arises: the sign of the
right-hand-side depends on the value $z_i$. This is easy to see, using the
fact that by symmetry $\rho_L(\theta)=\rho_R(-\t)$. Thus we must replace the
$\pm$ by $sgn(z_i)$. The TBA relations can then be derived, but many features
are more complicated. The kernels $\phi_{ab}$ involve this irritating $sgn$,
and are not symmetric in $a$ and $b$. As a result, the densities one derives
have discontinuities. However, the TBA analysis can still be formally done,
although we are not convinced that it means anything. In particular, when one
tries to add fugacties to truncate to the minimal models, the equations
produce a complex answer even though the energy should remain real. In any
case, we have studied the TBA equations for this case $\mu=1/2$. We obtain an
effective central charge which starts at $.2$ in the UV and increases smoothly
and monotonically to $1$ in the IR. The tricks one usually uses to rewrite the
ground-state energy in the ultraviolet limit as a sum of dilogarithms must be
modified because of the discontinuities, but they give the result of $.2$.

In the general case $\mu\ne\half$ outside of the IR limit, things are even
more complicated. The problems with signs are worse, and it is not even clear
how to classify the solutions of the equation \fory. The usual string
hypothesis \rTS\ is not valid, because it destroys the left-right symmetry. We
have made a modified string hypothesis for $1/\mu$ an integer, where there
is an antistring for every string. This restores the left-right symmetry. We
have studied the resulting TBA systems, but not found anything of interest
yet.  Again, it is not clear whether this has any meaning. The fact that these
extra antistrings must disappear in the IR limit where we understand the
system in fact leads us to believe that the solutions of these TBA equations
may not be simply described in terms of strings.

\listrefs
\end